\documentclass[a4paper,fleqn,usenatbib]{mnras}
\usepackage[dvipsnames,usenames]{color}
\usepackage{colortbl}
\usepackage{mathptmx}
\usepackage{etoolbox}
\usepackage{graphicx}
\usepackage{amsmath}
\usepackage{natbib}

\newcommand{\etal}{{et al}\/.}

\pubyear{2019}
\begin{document}
\title[NGC 326]{NGC 326: X-shaped no more}
\author[M.J.\ Hardcastle et al.]
       {M.J.\ Hardcastle$^1$, J.H.\ Croston$^2$, T.W.\ Shimwell$^{3,4}$,
         C.\ Tasse$^{5,6}$,
  G.\ G\"urkan$^7$, 
  \newauthor R. Morganti$^{3,8}$, M.\ Murgia$^9$, H.J.A.\ R\"ottgering$^4$, R.J. van Weeren$^4$ and W.L.\ Williams$^4$\\
$^1$ Centre for Astrophysics Research, School of Physics, Astronomy and Mathematics, University of
  Hertfordshire, College Lane, Hatfield AL10 9AB, UK\\
$^2$  Department of Physical Sciences, The Open University, Milton
  Keynes MK7 6AA, UK\\
$^3$ ASTRON, Netherlands Institute for Radio Astronomy, Oude
  Hoogeveensedijk 4, Dwingeloo, 7991 PD, The Netherlands\\
$^4$ Leiden Observatory, Leiden University, PO Box 9513, 2300 RA Leiden, The Netherlands\\
$^5$ GEPI, Observatoire de Paris, CNRS, Universit\'e Paris Diderot, 5
  place Jules Janssen, 92190 Meudon, France\\
$^6$ Department of Physics \& Electronics, Rhodes University, PO Box 94, Grahamstown, 6140, South Africa\\
$^7$ CSIRO Astronomy and Space Science, PO Box 1130 \& Bentley WA
  6102, Perth, Australia\\
$^8$  Kapteyn Astronomical Institute, University of Groningen, P.O. Box 800,
9700 AV Groningen, The Netherlands\\
$^9$  INAF -- Osservatorio Astronomico di Cagliari, Via della Scienza
  5, I-09047 Selargius (CA), Italy\\
}
\maketitle
\begin{abstract}
We present new 144-MHz LOFAR observations of the prototypical
`X-shaped' radio galaxy NGC 326, which show that the formerly known
wings of the radio lobes extend smoothly into a large-scale, complex
radio structure. We argue that this structure is most likely the result of
hydrodynamical effects in an ongoing group or cluster merger, for
  which pre-existing X-ray and optical data provide independent evidence. The
  large-scale radio structure is hard to explain purely in terms of
  jet reorientation due to the merger of binary black holes, a
  previously proposed explanation for the inner structure of NGC 326.
  For this reason, we suggest that the simplest model is one in which
  the merger-related hydrodynamical processes account
  for all the source structure, though we do not rule out the
  possibility that a black hole merger has occurred. Inference of the
black hole-black hole merger rate from observations of X-shaped
sources should be carried out with caution in the absence of deep,
sensitive low-frequency observations. Some X-shaped sources may be
signposts of cluster merger activity, and it would be useful to
investigate the environments of these objects more generally.
\end{abstract}
\begin{keywords}
galaxies: jets -- galaxies: active -- radio continuum: galaxies
\end{keywords}

\section{Introduction}
\label{sec:intro}

As every massive galaxy contains a supermassive black hole (BH:
\citealt{Magorrian+98}), the formation of binary systems of black
holes is widely believed to be the inevitable consequence of the
observed major mergers between massive galaxies \citep{Begelman+80}.
It remains unclear theoretically, as noted by \citeauthor{Begelman+80},
whether the binary pairs thus formed can merge on a timescale shorter
than the Hubble time, though recent work suggests that this is indeed
possible \citep[e.g][]{Gualandris+17}. Constraints on the supermassive
BH merger rate are important not just because they constrain
cosmological models of galaxy formation and evolution, but also because
they provide predictions for the rates of BH-BH merging events
\citep{Wyithe+03} in future gravitational wave detectors such as
pulsar timing arrays \citep{Hobbs+10} or LISA \citep{Amaro-Seoane+12}.

Radio-loud active galaxies (RLAGN) -- radio galaxies and radio-loud
quasars -- are important in studies of the supermassive BH merger rate
because the synchrotron-emitting plasma deposited by their jets gives
us a fossil record of the jet orientation over timescales of the radio
source lifetime (which may be hundreds of Myr or more). Assuming, as
is widely accepted, that the axis of jet generation is determined by
the black hole spin axis \citep{Blandford+Znajek77}, then the
formation of a close binary BH of which one member is a RLAGN will
lead first to signatures of jet precession \citep[e.g.][]{Krause+19}
followed by, eventually, an abrupt re-orientation of the jet axis as
the two BH merge into one. Of course, this scheme assumes that the
conditions for jet generation (non-negligible rates of accretion of
magnetized material) can persist during the close binary phase and be
re-established after merger, which may not be the case. Nevertheless,
it is important to search for examples of RLAGN that provide evidence
for this merger process.

In an influential paper \cite{Merritt+Ekers02} argued that the
X-shaped radio sources provide direct evidence for BH-BH mergers.
These are RLAGN which show a pair of extended `wings' at a large angle
to the currently active pair of lobes. In the BH-BH merger model, the
wings represent the former lobe direction before jet reorientation,
while the current lobes tell us about the current jet axis.
Hydrodynamical models, in which the wings are simply distorted
backflow from the active lobes \citep{Leahy+Williams84}, have
difficulties in explaining systems in which the wings are longer than
the active lobes without appealing to a peculiar source environment,
and so the merger explanation for the X-shaped source class is
attractive, and has motivated a number of searches for evidence of
BH-BH mergers or binary BH in these systems. Samples of candidate
X-shaped galaxies have been generated by visual inspection of existing
radio catalogues in order to carry out such searches
\citep{Cheung+07b,Cheung+09}, and tests of expectations of the
different models have been carried out, with some observations tending
to favour unusual environments for X-shaped sources
\citep{Hodges-Kluck+10,Landt+10} while others have argued for a binary black-hole
origin \citep{Zhang+07}.

The first winged source to be discovered \citep{Ekers+78}, and a
prototype of the X-shaped class \citep{Wirth+82}, is the nearby radio
galaxy B2 0055+26 or 4C\,26.03, usually referred to by the name of its optical
identification, NGC 326. In this paper we report on new observations of this
galaxy with the Low-Frequency Array (LOFAR: \citealt{vanHaarlem+13})
that show conclusively that the large-scale structure of the source is
generated by hydrodynamical effects, presumably related to bulk motion
with respect to the ambient medium. We argue that high-quality,
sensitive observations are necessary before coming to the conclusion
that any particular radio morphology is indicative of BH-BH merger.

Throughout the paper we use a cosmology in which $H_0 = 70$ km
s$^{-1}$, $\Omega_{\rm m} = 0.3$ and $\Omega_\Lambda = 0.7$. At the
redshift of NGC 326 1 arcsec corresponds to 0.93 kpc. The
spectral index $\alpha$ is defined in the sense $S \propto
\nu^{-\alpha}$.

\section{NGC 326}

\begin{figure*}
  \includegraphics[width=1.0\linewidth]{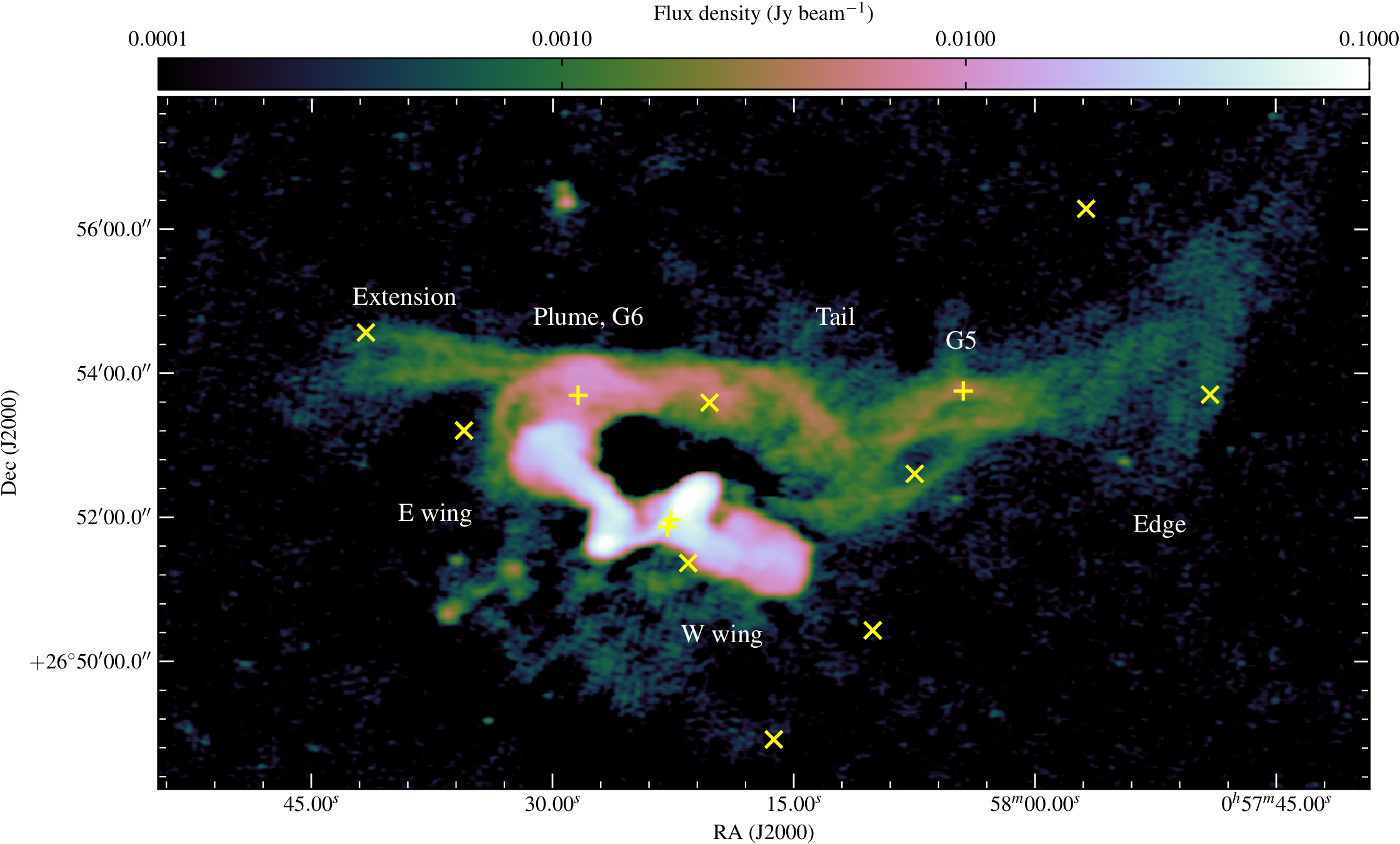}
  \caption{LOFAR colourscale of the radio emission from NGC 326. The
    colour scale is logarithmic over the range shown by the colour bar
    (in units of Jy beam$^{-1}$ for the resolution of $8.2 \times 5.1$ arcsec).
    yellow crosses mark the positions of potential cluster members;
    $+$ signs are those in the spectroscopic study of
    \protect\cite{Werner+99} and $\times$ signs show candidate galaxies selected
    from NED as described in the text. The double nuclei of NGC 326
    are represented by the two adjacent crosses close to the
    centre of the image. Features referred to in the text are labelled.}
  \label{fig:lofar}
\end{figure*}

B2 0055+26 is a radio galaxy at a redshift of 0.0474 \citep{Werner+99}.
Its total radio luminosity of $8 \times 10^{24}$ W Hz$^{-1}$ at 1.4
GHz \citep{Murgia+01} places it just below the nominal Fanaroff-Riley
luminosity break \citep{Fanaroff+Riley74} which tends to separate
sources with and without compact hotspots, in contrast to other
well-studied X-shaped sources which are generally FRII radio sources.
The best radio images to date are those of \cite{Murgia+01}, who
observed it at several frequencies with the (pre-upgrade) NRAO Very Large
Array (VLA), and their detailed total intensity and polarization maps are
consistent with this picture; the source shows the centre-brightened,
transverse-magnetic-field jets of an FRI source and little compact
structure in the extended lobes.

The host galaxy, NGC 326 itself, is a dumb-bell galaxy with two
optical nuclei \citep{Wirth+82}, with a projected separation of 6.6
kpc. Both nuclei are radio sources \citep{Murgia+01} with the jets
being associated with the northern one (Core 1 in the notation of
\citealt{Werner+99}). The host galaxy is the brightest member of a
small optical group, Zwicky 0056.9+2626 \citep{Zwicky+Kowal68}.
Optical spectroscopy by \cite{Werner+99} measured a velocity
dispersion of $599^{+230}_{-107}$ km s$^{-1}$, consistent with a poor
cluster. In the X-ray, the environment was studied with the {\it
  ROSAT} PSPC by \cite{Worrall+95}, who found asymmetrical diffuse
X-ray emission with a temperature $\sim 2$ keV, consistent with the
velocity dispersion in implying a poor cluster environment; the X-ray
environment is bright enough, at the low redshift of the system, that
the cluster is a member of the {\it ROSAT} Brightest Cluster Sample of
\cite{Ebeling+98} under the name RXJ0058.9+2657. More recently
\cite{Hodges-Kluck+Reynolds12} presented a {\it Chandra} observation
which confirms the asymmetrical nature of the X-rays.

\section{Observations}

\subsection{LOFAR}

\begin{table}
  \caption{Observations of NGC326 with LOFAR}
  \label{tab:observations}
  \begin{tabular}{llrr}
    \hline
    Field ID&Observation date&Duration (h)&Target offset from\\
    &&&pointing centre (deg.)\\
    \hline
    P013+26&2016-09-19&8&1.43\\
    P013+26&2017-01-16&8&1.43\\
    P016+26&2016-10-14&8&1.65\\
    P014+29&2016-10-03&8&1.68\\
    \hline
  \end{tabular}
  \end{table}

LOFAR has observed the NGC 326 field as part of the LOFAR Two-metre
Sky Survey, LoTSS\footnote{\url{https://lofar-surveys.org/}}, a deep
survey of the northern sky at 144 MHz \citep{Shimwell+17}. Four
pointings were used to construct the image used in the current paper:
observational details are listed in Table \ref{tab:observations}. The
data were initially reduced with version 2.2 of the standard Surveys
Key Science Project
pipeline\footnote{\url{https://github.com/mhardcastle/ddf-pipeline}},
as described by \cite{Shimwell+19}. This pipeline carries out
direction-dependent calibration using {\sc killMS}
\citep{Tasse14,Smirnov+Tasse15} and imaging is done using {\sc
  DDFacet} \citep{Tasse+18}. Version 2.2 of the pipeline makes use of
enhancements to the calibration and imaging, particularly of extended
sources, that were described briefly in section 5 of
\cite{Shimwell+19} and will be discussed more fully by Tasse
\etal\ (in prep.). Running this pipeline gives us a mosaiced image of
the field around NGC326, in which the images made from the three
separate pointings (P013+26 was observed twice in error) are combined
in the image plane, with a resolution of 6 arcsec and an rms noise of
95 $\mu$Jy beam$^{-1}$.

The surveys pipeline finds self-calibration solutions for large areas
of sky, and is not expected necessarily to find the optimal solution
for any given sky position. To enhance the quality of the images
around NGC 326 further, we used the models derived from the pipeline
to subtract off all modelled sources from all four fields, leaving
only the data for NGC 326 and its immediate surroundings, averaged
appropriately, and then carried out several iterations of phase and
amplitude self-calibration on the combined dataset (after correction
for the LOFAR station beam at the source position) in order to improve
the accuracy of the calibration solutions at that location, using {\sc
  wsclean} \citep{Offringa+14} as the imager. This process (which will
be described in more detail by van Weeren \etal\ in prep.) gave us an
image at the central frequency of 144 MHz with a resolution of $8.2
\times 5.1$ arcsec (beam position angle of $84^\circ$ and an rms noise
level around the radio source of 110 $\mu$Jy beam$^{-1}$. The image
noise is slightly worse in this post-processed image, presumably as a
result of residuals from the subtraction being averaged over the
image, but the image fidelity should be better since the pointings are
combined in the $uv$ plane rather than the image plane and since we
can derive accurate amplitude and phase solutions at the source
position. We therefore adopt the post-processed image for our detailed
study.

The LOFAR flux scale is known to be significantly uncertain due to
difficulties in transferring calibration from a reference source to
the target field, and the bootstrap process applied to the data
\citep{Hardcastle+16} does not necessarily completely counteract
either this or self-calibration-induced flux-scale drift. The
integrated 144-MHz flux density of NGC 326 ($9.96 \pm 0.01$ Jy)
compares well with the Culgoora 160-MHz flux density of 9.5 Jy
\citep{Slee95}, given that the systematic errors on both flux
densities are of order 10 per cent. The 4C flux density of $5.1 \pm 0.6$ Jy
at 178 MHz \citep{Pilkington+Scott65} will have been severely affected
by resolution by the 4C interferometer and is not reliable. We were
also able to check the flux scale with reference to the nearby bright
(and relatively compact) source 3C\,28, which has a measured 144-MHz
flux density on the mosaiced pipeline output of 21.4 Jy, compared to
an expectation from the measurements given by \cite{Laing+83} of 22.3
Jy (probably again with a $\sim 10$ per cent uncertainty). We conclude
that the flux scale of the image is adequate for the purposes of this
paper without further correction.

\subsection{X-ray data}

We downloaded the {\it Chandra} dataset, obsid 6830, described by
\cite{Hodges-Kluck+Reynolds12} from the public archive and
reprocessed it using {\sc ciao} 4.7 and the latest {\it Chandra}
processing threads. In addition, we downloaded the {\it ROSAT} PSPC
total-band image and exposure maps from the HEASARC archive and made a
crude exposure-corrected image with 15-arcsec pixels, using the same data as described by
\cite{Worrall+95}, to illustrate the extent of the large-scale X-ray
emission. As the focus of this work is the large-scale structure we
have not applied any astrometric correction to the {\it ROSAT} data.

\subsection{Optical data}

For the purposes of probing the large-scale optical environment we
supplemented the galaxies described by \cite{Werner+99} with other
galaxies with spectroscopic redshifts taken from the Nasa
Extragalactic Database (NED): principal redshift sources for NED
include the Sloan Digital Sky Survey and the spectroscopic survey of
\cite{Cava+09}, who identified 22 cluster member candidates.
Here we take as cluster member candidates any NED galaxy within 30 arcmin
(1.7 Mpc) of NGC 326 having a velocity within
1500 km s$^{-1}$ of the central velocity of 14,307 km s$^{-1}$ given
by \cite{Werner+99} --- the velocity range here is approximately 2.5
times the velocity dispersion of \citeauthor{Werner+99}. This
gives a sample of 32 galaxies (including both nuclei of NGC 326) with
a mean velocity of 14,530 km s$^{-1}$ and a velocity disperson of 630
km s$^{-1}$, consistent with the measurements of
\citeauthor{Werner+99}. The list of cluster member candidates is given
in Table \ref{tab:galaxies}.

\begin{table*}
  \caption{Galaxies selected from NED within 30 arcmin of NGC 326 and
    close in velocity space as described in the text. Data taken from
    NED.}
  \label{tab:galaxies}
  \begin{tabular}{lllrrl}
    \hline
    Source name&RA&Dec&Heliocentric velocity (km s$^{-1}$)&2MASS $K_s$&Werner name\\
    \hline
NGC 0326 1&00h58m22.6s&+26d51m58.5s &$14610 \pm 15$ &11.822&G1-1\\
NGC 0326 2&00h58m22.8s&+26d51m52.4s &$14822 \pm 25$ &11.971&G1-2\\
2MASS J00582156+2651219&00h58m21.6s&+26d51m22.3s &$13996 \pm 7$ &13.658\\
2MASS J00582023+2653357&00h58m20.2s&+26d53m35.9s &$13809 \pm 27$ &14.820\\
2MASX J00582840+2653420&00h58m28.4s&+26d53m42.0s &$15589 \pm 30$ &12.742&G6\\
2MASX J00583550+2653121&00h58m35.5s&+26d53m12.6s &$15600 \pm 5$ &13.957\\
2MASS J00581008+2650251&00h58m10.1s&+26d50m26.2s &$14329 \pm 31$ &14.940\\
2MASS J00581623+2648549&00h58m16.2s&+26d48m55.4s &$13737 \pm 17$ &15.383\\
GALEXASC J005807.41+265235.5&00h58m07.5s&+26d52m36.6s &$14351 \pm 87$ &No 2MASS\\
MGC +04-03-024&00h58m04.5s&+26d53m45.6s &$13850 \pm 60$ &13.539&G5\\
2MASX J00580956+2647596&00h58m09.6s&+26d47m59.6s &$14000 \pm 30$ &12.680&G8\\
WINGS J005841.63+265434.1&00h58m41.6s&+26d54m34.1s &$15795 \pm 88$ &No photometry\\
2MASX J00580702+2647545&00h58m07.0s&+26d47m55.5s &$14776 \pm 22$ &13.960\\
2MASS J00575680+2656170&00h57m56.8s&+26d56m17.3s &$15753 \pm 19$ &14.691\\
2MASS J00574910+2653421&00h57m49.1s&+26d53m42.5s &$13593 \pm 130$ &$>14.939$\\
WINGS J005812.33+264355.7&00h58m12.3s&+26d43m55.7s &$14497 \pm 57$ &No photometry\\
2MASS J00584743+2658393&00h58m47.4s&+26d58m39.3s &$13519 \pm 19$ &12.277&G4\\
2MASX J00591642+2652009&00h59m16.4s&+26d52m01.7s &$14856 \pm 25$ &12.880\\
2MASS J00591141+2658121&00h59m11.4s&+26d58m12.4s &$14517 \pm 99$ &No 2MASS\\
2MASX J00572524+2650103&00h57m25.3s&+26d50m10.3s &$14901 \pm 14$ &12.846\\
MCG +04-03-030&00h59m03.6s&+27d02m32.8s &$14570 \pm 30$ &12.260&G3\\
2MASX J00592575+2647599&00h59m25.8s&+26d48m00.7s &$14841 \pm 43$ &13.428\\
2MASX J00592914+2655299&00h59m29.2s&+26d55m30.1s &$14833 \pm 43$ &13.727\\
2MASX J00593281+2649476&00h59m32.8s&+26d49m48.0s &$13875 \pm 34$ &14.320\\
2MASX J00591793+2704069&00h59m17.9s&+27d04m06.7s &$15518 \pm 75$ &13.456\\
2MASX J00585376+2708030&00h58m53.8s&+27d08m03.7s &$14901 \pm 25$ &13.327\\
UGC 613&00h59m24.4s&+27d03m32.6s &$13770 \pm 23$ &11.894&G2\\
2MASX J00590040+2708469&00h59m00.4s&+27d08m46.8s &$14390 \pm 30$ &12.139&G7\\
2MASXi J0059248+270720&00h59m24.8s&+27d07m21.0s &$14386 \pm 66$ &14.466\\
2MASX J00594231+2639026&00h59m42.3s&+26d39m02.8s &$14145 \pm 45$ &13.681\\
UGC 585&00h56m45.4s&+27d00m34.8s &$13938 \pm 22$ &12.075\\
CGCG 480-031&01h00m27.9s&+27d01m30.8s &$14914 \pm 16$ &13.470\\
\hline
  \end{tabular}
  \end{table*}

\section{Images}

\begin{figure*}
  \includegraphics[width=1.0\linewidth]{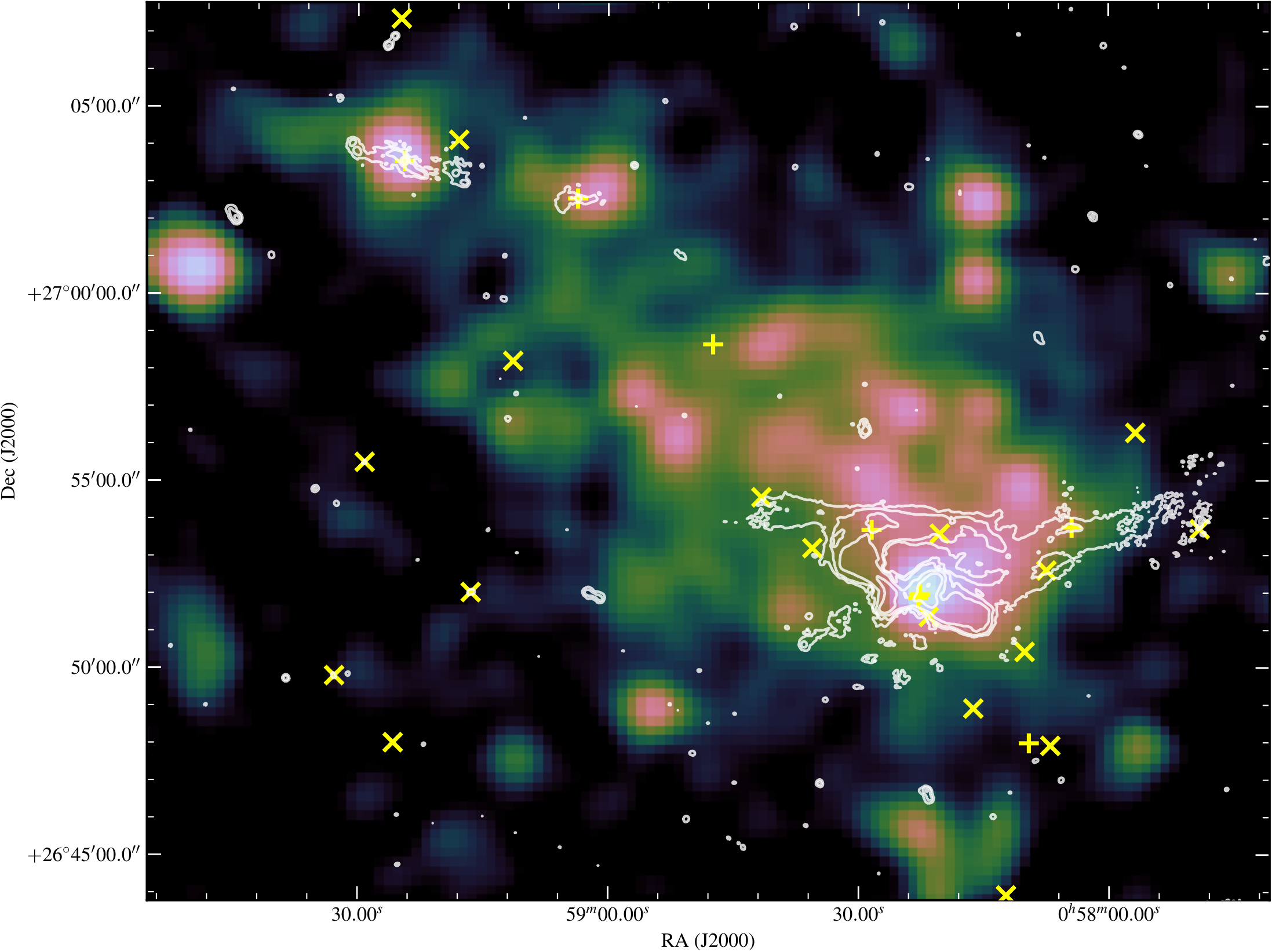}
  \caption{LOFAR contours of radio emission from the NGC 326 field
    (white, at $0.5 \times (1, 4, 16\dots)$ mJy beam$^{-1}$)
    overlaid on a Gaussian-smoothed exposure-corrected {\it ROSAT}
    image with a logarithmic transfer function (colour). Crosses as in Fig. \ref{fig:lofar}.}
  \label{fig:rosat}
\end{figure*}

Fig.\ \ref{fig:lofar} shows the LOFAR image of NGC 326 as a
colourscale. This image has several interesting features. The X-shaped
structure seen in all previous observations is clearly visible in the
centre of the image and is structurally very similar to what was seen
in the images of \cite{Murgia+01}. However, the east wing of the
source has a clear and relatively bright extension, albeit with a drop
of a factor $\sim 3$ in surface brightness, which continues northwards
(the `plume' just visible in the images of \citeauthor{Murgia+01})
before bending sharply westward into a filamentary `tail' structure
that extends a total of 13 arcmin (720 kpc in projection) in an
east-west direction. This feature, because of its angular size, would
have been largely invisible to the VLA C-configuration L-band
observations of \citeauthor{Murgia+01} which were only sensitive to
structures of scale $<6$ arcmin. If they had had the required short
baselines, the surface brightness of 4 mJy beam$^{-1}$ at 144 MHz in
the bright parts of the tail should have been detectable to them at
their observing frequency for $\alpha \sim 1$. No sensitive VLA
D-configuration L-band observations of the target exist and so we
cannot determine the LOFAR to VLA spectral index. Parts of the tail
are plausibly detected at low significance in the NVSS survey, with
the benefit of the LOFAR images to guide the eye, but are not well
enough resolved from the wings and lobes or of sufficient signal to
noise to allow a spectral index measurement. The integrated flux
measurements of \cite{Slee95} imply $\alpha \approx 1.0$ for the whole
source between 80 and 160 MHz and so it is plausible that the `tails'
are steep-spectrum.

A few other features of the tail structure are noteworthy. The
connection between the west wing of the source and the tail is not as
obvious as for the east wing, but several filamentary structures
extend from the west wing towards the tail and may merge with it. One
possibility is that there would be a high-surface-brightness extension
of the west wing that is bent back behind the wing itself at our
viewing angle; the wings may be much longer in projection than they
are on the sky. In any case it seems likely that the tail structure is
composed of plasma that originated in both lobes/wings of the radio
source; whether these are merged or just superposed in projection is
not clear.

The filamentary structures in the tail are very striking. They cannot
be imaging/deconvolution artefacts, since they appear essentially
identical in our two high-resolution images made using {\sc wsclean}
and {\sc ddfacet}. We believe they provide evidence for large-scale
coherent magnetic field enhancements: since we expect that the
particles dominate the energy density in the tails \citep{Croston+18},
field enhancements can exist and persist without creating a
significant pressure difference within the tails. Similar filamentary
structures exist in LOFAR observations of other tailed sources (e.g.
3C\,83.1B, Bempong-Manful et al in prep.) -- the excellent $uv$ plane
coverage of full-synthesis LOFAR observations makes them particularly
sensitive to such structures. There is no immediately obvious
explanation for the filamentary extension of the tail to the E of
where it is entered by the plume of the E tail, or for its relatively
sharp edge at the W end (see labels in Fig.\ \ref{fig:lofar} for the
locations of these features).

Finally, we notice the remarkable alignment of 7 cluster candidate
galaxies in an E-W direction on or close to the tail. The galaxy 2MASX
J00582840+2653420 (G6 in the notation of \citealt{Werner+99}), which
as noted by \cite{Murgia+01} is superposed on the `plume', is not
associated with any radio structure in the image; nor are any of the
others, with the exception of MCG +04-03-024 (Werner's G5, 2MASX J00580439+2653455),
which lies directly on top of one of the bright filaments in the tail and has
a compact radio counterpart. This galaxy is a disky system in the
POSS-II images supplied by NED and has a bright X-ray counterpart in the {\it Chandra}
data, probably making it a Seyfert galaxy; it does not seem likely to
us that it is directly responsible for any of the large-scale
emission, including the filament it lies on, but clearly we cannot
rule this out.

Fig.\ \ref{fig:rosat} shows the LOFAR image overlaid on the {\it
  ROSAT} data originally presented by \cite{Worrall+95}. As they
noted, the X-ray emission shows a strong elongation in a NE-SW
direction, with NGC 326 lying to the SW. Another galaxy in the group,
UGC 613 (G2 of \citealt{Werner+99}) is seen in the LOFAR images to be
an extended radio source, with a flux density of around 300 mJy at 144
MHz, and has a compact {\it ROSAT} counterpart. A fainter extended
radio source is associated with the neighbouring galaxy MCG +04-03-030, and
this also has a much weaker X-ray counterpart seen in the {\it ROSAT}
and {\it Chandra} data. Other radio sources in the field appear to be
unrelated to the group. The X-ray properties of the group, together
with the roughly bimodal distribution of member galaxies on the sky
(one sub-group concentrated around NGC 326 and one more north-south
extended group with UGC 613 as its most prominent member) are strongly
suggestive of an unrelaxed, merging system. There is no clear
relationship between the E-W direction of the radio tail and the NE-SW
extension of the X-rays, either in the {\it ROSAT} or the
higher-resolution {\it Chandra} data.

\section{Discussion and conclusions}

The most obvious conclusion to be drawn from the deep LOFAR image is
that NGC 326 can no longer be viewed as an archetype of the model in
which jets are suddently reoriented by BH-BH merger. The argument
  in favour of BH-BH merger for this source \citep{Merritt+Ekers02}
  was based on the idea that hydrodynamical explanations for the
  lobe-wing structure, requiring complex hydrodynamic structure in the
  intra-cluster medium, were intrinsically less plausible than BH-BH
  mergers, which must after all occur. However, the extension of the
wings into the large-scale tail structure {\it requires} a
hydrodynamical explanation that can give rise to apparent sharp
  bends in the radio structure, and so in turn disfavours (on
  the principle of seeking the simplest possible explanation for the
  phenomena) any other explanation for the generation of the wings
themselves. We emphasise that a BH-BH merger, either manifesting
  as an abrupt jet reorientation or as a slower transition mediated by
  gas associated with the merger as proposed by \cite{Zier05}, is not
ruled out by these data, but it is no longer either a necessary or a
sufficient explanation for the observed source morphology on its own.
\cite{Worrall+95}, who favoured buoyancy as the bending mechanism for
the wings, suggested that the extension of the E wing that was just
visible in their data might be due to the plasma from the wing
reaching neutral buoyancy (i.e. the point where the density of the
radio structure matches that of its surroundings), but it is difficult
to see that on its own this model can account for the scale of the
tail structure that we now see, with an elongation several hundred kpc
away from the injection point through what are presumably very
different conditions in the intracluster medium. While buoyancy must
necessarily play a role, we suggest that complex large-scale bulk
motions within the X-ray-emitting medium induced by the ongoing
cluster merger, coupled with motions of the
host galaxy itself with respect to that medium, are the only viable
explanation for the observed radio structure in NGC 326. The same
  complex cluster hydrodynamics which account for the tail can
  plausibly then also account for the bending of the (no doubt
  projected) inner lobes into the wings. Such explanations still face
difficulties; the 700-kpc projected length of the tail, if generated
in a plausible AGN lifetime of $10^8$ years, requires bulk growth
speeds around 7000 km s$^{-1}$, much higher than the velocity
dispersion or sound speed of the cluster, and so either high bulk
speeds or a much larger source age are required. The significant
change in surface brightness of the radio structures between the wings
and the tail may also suggest some intermittency in the energy supply.
Spectral index studies of NGC 326 will shed light on its history: the
source has been observed at lower frequencies with the LOFAR LBA, and
the new observations, together with complementary VLA and GMRT
observations, will be used to study the spectral and other properties
of the newly detected features in a forthcoming paper (Murgia et al.
in prep.).

In the model we prefer NGC 326 becomes a member of the growing
class of objects exhibiting this kind of complex interaction between
the AGN-injected cosmic ray electrons and the cluster gas (see e.g.
\citealt{vanWeeren+19} for a review). As in some of the best-studied
cluster systems, there is no real boundary between plasma associated
with the radio galaxy and material that is presumably moving with, and
plausibly thoroughly mixed with, the intracluster medium; when the
jets of NGC 326 switch off, the result will be a merging cluster with
a population of distributed energetic cosmic rays and associated
magnetic fields which may be re-energised by later shocks or
compression to give rise to diffuse radio emission. Observations that
allow the measurement of the spectral index or of polarization in
the tail will help us to assess the extent to which it is now
responding to the intracluster medium. There is little or no
significantly detected polarization in an RM synthesis analysis of the
LOFAR HBA data.

Observers inferring the presence of binary BH or BH-BH merger from
X-shaped radio sources should be alert to the possibility of
alternative, hydrodynamical explanations for the radio morphology and
should ideally, before subscribing to a merger explanation, check that
the wings in the sources of interest are sharply bounded in deep radio
observations with good sensitivity to large angular scales. The LoTSS
survey will allow the study of many more X-shaped sources in detail in
the coming years.

\section*{Acknowledgments}

MJH and JHC acknowledge support from the UK Science and Technology
Facilities Council ([ST/R000905/1], [ST/R00109X/1] and
[ST/R000794/1]). HR and WLW acknowledge support from the ERC Advanced
Investigator programme NewClusters 321271. RJvW acknowledges support
of the VIDI research programme with project number 639.042.729, which
is financed by the Netherlands Organisation for Scientific Research
(NWO). We thank A.\ Shulevski, M.G.H.\ Krause, S.\ O'Sullivan, and
E.\ Brinks for helpful comments on an earlier draft,
and an anonymous referee for a constructive reading of the paper.

This research has made use of the University
of Hertfordshire high-performance computing facility
(\url{https://uhhpc.herts.ac.uk/}) and the LOFAR-UK compute facility,
located at the University of Hertfordshire and supported by STFC
[ST/P000096/1].. This research made use of {\sc Astropy}, a
community-developed core Python package for astronomy
\citep{AstropyCollaboration13} hosted at
\url{http://www.astropy.org/}, and of {\sc APLpy}, an open-source
plotting package for Python \citep{Robitaille+Bressert12}. This
research has made use of the NASA/IPAC Extragalactic Database (NED),
which is operated by the Jet Propulsion Laboratory, California
Institute of Technology, under contract with the National Aeronautics
and Space Administration.

LOFAR, the Low Frequency Array, designed and constructed by ASTRON, has
facilities in several countries, which are owned by various parties
(each with their own funding sources), and are collectively operated
by the International LOFAR Telescope (ILT) foundation under a joint
scientific policy. The ILT resources have benefited from the
following recent major funding sources: CNRS-INSU, Observatoire de
Paris and Universit\'e d'Orl\'eans, France; BMBF, MIWF-NRW, MPG, Germany;
Science Foundation Ireland (SFI), Department of Business, Enterprise
and Innovation (DBEI), Ireland; NWO, The Netherlands; the Science and
Technology Facilities Council, UK; Ministry of Science and Higher
Education, Poland.

\bibliographystyle{mnras}
\renewcommand{\refname}{REFERENCES}
\bibliography{../bib/new-temp,../bib/mjh,../bib/cards}

\begin{thebibliography}{}
\makeatletter
\relax
\def\mn@urlcharsother{\let\do\@makeother \do\$\do\&\do\#\do\^\do\_\do\%\do\~}
\def\mn@doi{\begingroup\mn@urlcharsother \@ifnextchar [ {\mn@doi@}
  {\mn@doi@[]}}
\def\mn@doi@[#1]#2{\def\@tempa{#1}\ifx\@tempa\@empty \href
  {http://dx.doi.org/#2} {doi:#2}\else \href {http://dx.doi.org/#2} {#1}\fi
  \endgroup}
\def\mn@eprint#1#2{\mn@eprint@#1:#2::\@nil}
\def\mn@eprint@arXiv#1{\href {http://arxiv.org/abs/#1} {{\tt arXiv:#1}}}
\def\mn@eprint@dblp#1{\href {http://dblp.uni-trier.de/rec/bibtex/#1.xml}
  {dblp:#1}}
\def\mn@eprint@#1:#2:#3:#4\@nil{\def\@tempa {#1}\def\@tempb {#2}\def\@tempc
  {#3}\ifx \@tempc \@empty \let \@tempc \@tempb \let \@tempb \@tempa \fi \ifx
  \@tempb \@empty \def\@tempb {arXiv}\fi \@ifundefined
  {mn@eprint@\@tempb}{\@tempb:\@tempc}{\expandafter \expandafter \csname
  mn@eprint@\@tempb\endcsname \expandafter{\@tempc}}}

\bibitem[\protect\citeauthoryear{{Amaro-Seoane} et~al.,}{{Amaro-Seoane}
  et~al.}{2012}]{Amaro-Seoane+12}
{Amaro-Seoane} P.,  et~al., 2012, \mn@doi [Classical and Quantum Gravity]
  {10.1088/0264-9381/29/12/124016}, \href
  {http://adsabs.harvard.edu/abs/2012CQGra..29l4016A} {29, 124016}

\bibitem[\protect\citeauthoryear{{Astropy Collaboration} et~al.,}{{Astropy
  Collaboration} et~al.}{2013}]{AstropyCollaboration13}
{Astropy Collaboration} et~al., 2013, \mn@doi [\aap]
  {10.1051/0004-6361/201322068}, \href
  {http://adsabs.harvard.edu/abs/2013A%26A...558A..33A} {558, A33}

\bibitem[\protect\citeauthoryear{{Begelman}, {Blandford}  \& {Rees}}{{Begelman}
  et~al.}{1980}]{Begelman+80}
{Begelman} M.~C.,  {Blandford} R.~D.,   {Rees} M.~J.,  1980, \mn@doi [\nat]
  {10.1038/287307a0}, \href
  {https://ui.adsabs.harvard.edu/\#abs/1980Natur.287..307B} {287, 307}

\bibitem[\protect\citeauthoryear{{Blandford} \& {Znajek}}{{Blandford} \&
  {Znajek}}{1977}]{Blandford+Znajek77}
{Blandford} R.~D.,  {Znajek} R.~L.,  1977, \mn@doi [\mnras]
  {10.1093/mnras/179.3.433}, \href
  {http://adsabs.harvard.edu/abs/1977MNRAS.179..433B} {179, 433}

\bibitem[\protect\citeauthoryear{{Cava} et~al.,}{{Cava} et~al.}{2009}]{Cava+09}
{Cava} A.,  et~al., 2009, \mn@doi [\aap] {10.1051/0004-6361:200810997}, \href
  {https://ui.adsabs.harvard.edu/abs/2009A&A...495..707C} {495, 707}

\bibitem[\protect\citeauthoryear{{Cheung}}{{Cheung}}{2007}]{Cheung+07b}
{Cheung} C.~C.,  2007, \mn@doi [\aj] {10.1086/513095}, \href
  {http://adsabs.harvard.edu/abs/2007AJ....133.2097C} {133, 2097}

\bibitem[\protect\citeauthoryear{{Cheung}, {Healey}, {Landt}, {Verdoes Kleijn}
  \& {Jord{\'a}n}}{{Cheung} et~al.}{2009}]{Cheung+09}
{Cheung} C.~C.,  {Healey} S.~E.,  {Landt} H.,  {Verdoes Kleijn} G.,
  {Jord{\'a}n} A.,  2009, \mn@doi [\apjs] {10.1088/0067-0049/181/2/548}, \href
  {http://adsabs.harvard.edu/abs/2009ApJS..181..548C} {181, 548}

\bibitem[\protect\citeauthoryear{{Croston}, {Ineson}  \&
  {Hardcastle}}{{Croston} et~al.}{2018}]{Croston+18}
{Croston} J.~H.,  {Ineson} J.,   {Hardcastle} M.~J.,  2018, \mn@doi [\mnras]
  {10.1093/mnras/sty274}, \href
  {http://adsabs.harvard.edu/abs/2018MNRAS.476.1614C} {476, 1614}

\bibitem[\protect\citeauthoryear{{Ebeling}, {Edge}, {Bohringer}, {Allen},
  {Crawford}, {Fabian}, {Voges}  \& {Huchra}}{{Ebeling}
  et~al.}{1998}]{Ebeling+98}
{Ebeling} H.,  {Edge} A.~C.,  {Bohringer} H.,  {Allen} S.~W.,  {Crawford}
  C.~S.,  {Fabian} A.~C.,  {Voges} W.,   {Huchra} J.~P.,  1998, \mn@doi
  [\mnras] {10.1046/j.1365-8711.1998.01949.x}, \href
  {http://adsabs.harvard.edu/abs/1998MNRAS.301..881E} {301, 881}

\bibitem[\protect\citeauthoryear{{Ekers}, {Fanti}, {Lari}  \& {Parma}}{{Ekers}
  et~al.}{1978}]{Ekers+78}
{Ekers} R.~D.,  {Fanti} R.,  {Lari} C.,   {Parma} P.,  1978, \nat, 276, 588

\bibitem[\protect\citeauthoryear{{Fanaroff} \& {Riley}}{{Fanaroff} \&
  {Riley}}{1974}]{Fanaroff+Riley74}
{Fanaroff} B.~L.,  {Riley} J.~M.,  1974, \mnras, 167, 31P

\bibitem[\protect\citeauthoryear{{Gualandris}, {Read}, {Dehnen}  \&
  {Bortolas}}{{Gualandris} et~al.}{2017}]{Gualandris+17}
{Gualandris} A.,  {Read} J.~I.,  {Dehnen} W.,   {Bortolas} E.,  2017, \mn@doi
  [\mnras] {10.1093/mnras/stw2528}, \href
  {http://adsabs.harvard.edu/abs/2017MNRAS.464.2301G} {464, 2301}

\bibitem[\protect\citeauthoryear{{Hardcastle} et~al.,}{{Hardcastle}
  et~al.}{2016}]{Hardcastle+16}
{Hardcastle} M.~J.,  et~al., 2016, \mn@doi [\mnras] {10.1093/mnras/stw1763},
  \href {http://adsabs.harvard.edu/abs/2016MNRAS.462.1910H} {462, 1910}

\bibitem[\protect\citeauthoryear{{Hobbs} et~al.,}{{Hobbs}
  et~al.}{2010}]{Hobbs+10}
{Hobbs} G.,  et~al., 2010, \mn@doi [Classical and Quantum Gravity]
  {10.1088/0264-9381/27/8/084013}, \href
  {http://adsabs.harvard.edu/abs/2010CQGra..27h4013H} {27, 084013}

\bibitem[\protect\citeauthoryear{{Hodges-Kluck} \& {Reynolds}}{{Hodges-Kluck}
  \& {Reynolds}}{2012}]{Hodges-Kluck+Reynolds12}
{Hodges-Kluck} E.~J.,  {Reynolds} C.~S.,  2012, \mn@doi [\apj]
  {10.1088/0004-637X/746/2/167}, \href
  {https://ui.adsabs.harvard.edu/\#abs/2012ApJ...746..167H} {746, 167}

\bibitem[\protect\citeauthoryear{{Hodges-Kluck}, {Reynolds}, {Cheung}  \&
  {Miller}}{{Hodges-Kluck} et~al.}{2010}]{Hodges-Kluck+10}
{Hodges-Kluck} E.~J.,  {Reynolds} C.~S.,  {Cheung} C.~C.,   {Miller} M.~C.,
  2010, \mn@doi [\apj] {10.1088/0004-637X/710/2/1205}, \href
  {http://adsabs.harvard.edu/abs/2010ApJ...710.1205H} {710, 1205}

\bibitem[\protect\citeauthoryear{{Krause} et~al.,}{{Krause}
  et~al.}{2019}]{Krause+19}
{Krause} M. G.~H.,  et~al., 2019, \mn@doi [\mnras] {10.1093/mnras/sty2558},
  \href {https://ui.adsabs.harvard.edu/\#abs/2019MNRAS.482..240K} {482, 240}

\bibitem[\protect\citeauthoryear{{Laing}, {Riley}  \& {Longair}}{{Laing}
  et~al.}{1983}]{Laing+83}
{Laing} R.~A.,  {Riley} J.~M.,   {Longair} M.~S.,  1983, \mnras, 204, 151

\bibitem[\protect\citeauthoryear{{Landt}, {Cheung}  \& {Healey}}{{Landt}
  et~al.}{2010}]{Landt+10}
{Landt} H.,  {Cheung} C.~C.,   {Healey} S.~E.,  2010, \mn@doi [\mnras]
  {10.1111/j.1365-2966.2010.17183.x}, \href
  {http://adsabs.harvard.edu/abs/2010MNRAS.408.1103L} {408, 1103}

\bibitem[\protect\citeauthoryear{{Leahy} \& {Williams}}{{Leahy} \&
  {Williams}}{1984}]{Leahy+Williams84}
{Leahy} J.~P.,  {Williams} A.~G.,  1984, \mnras, 210, 929

\bibitem[\protect\citeauthoryear{{Magorrian} et~al.,}{{Magorrian}
  et~al.}{1998}]{Magorrian+98}
{Magorrian} J.,  et~al., 1998, \mn@doi [\aj] {10.1086/300353}, \href
  {http://adsabs.harvard.edu/abs/1998AJ....115.2285M} {115, 2285}

\bibitem[\protect\citeauthoryear{{Merritt} \& {Ekers}}{{Merritt} \&
  {Ekers}}{2002}]{Merritt+Ekers02}
{Merritt} D.,  {Ekers} R.~D.,  2002, \mn@doi [Science]
  {10.1126/science.1074688}, \href
  {https://ui.adsabs.harvard.edu/\#abs/2002Sci...297.1310M} {297, 1310}

\bibitem[\protect\citeauthoryear{{Murgia}, {Parma}, {de Ruiter}, {Bondi},
  {Ekers}, {Fanti}  \& {Fomalont}}{{Murgia} et~al.}{2001}]{Murgia+01}
{Murgia} M.,  {Parma} P.,  {de Ruiter} H.~R.,  {Bondi} M.,  {Ekers} R.~D.,
  {Fanti} R.,   {Fomalont} E.~B.,  2001, \mn@doi [\aap]
  {10.1051/0004-6361:20011436}, \href
  {https://ui.adsabs.harvard.edu/\#abs/2001A&A...380..102M} {380, 102}

\bibitem[\protect\citeauthoryear{{Offringa} et~al.,}{{Offringa}
  et~al.}{2014}]{Offringa+14}
{Offringa} A.~R.,  et~al., 2014, \mn@doi [\mnras] {10.1093/mnras/stu1368},
  \href {http://adsabs.harvard.edu/abs/2014MNRAS.444..606O} {444, 606}

\bibitem[\protect\citeauthoryear{{Pilkington} \& {Scott}}{{Pilkington} \&
  {Scott}}{1965}]{Pilkington+Scott65}
{Pilkington} J.~D.~H.,  {Scott} J.~F.,  1965, Memoirs of the Royal Astronomical
  Society, \href {https://ui.adsabs.harvard.edu/abs/1965MmRAS..69..183P} {69,
  183}

\bibitem[\protect\citeauthoryear{{Robitaille} \& {Bressert}}{{Robitaille} \&
  {Bressert}}{2012}]{Robitaille+Bressert12}
{Robitaille} T.,  {Bressert} E.,  2012, {APLpy: Astronomical Plotting Library
  in Python}, Astrophysics Source Code Library (\mn@eprint {ascl} {1208.017})

\bibitem[\protect\citeauthoryear{{Shimwell} et~al.,}{{Shimwell}
  et~al.}{2017}]{Shimwell+17}
{Shimwell} T.~W.,  et~al., 2017, \mn@doi [\aap] {10.1051/0004-6361/201629313},
  \href {http://ukads.nottingham.ac.uk/abs/2017A%26A...598A.104S} {598, A104}

\bibitem[\protect\citeauthoryear{{Shimwell} et~al.,}{{Shimwell}
  et~al.}{2019}]{Shimwell+19}
{Shimwell} T.~W.,  et~al., 2019, \mn@doi [\aap] {10.1051/0004-6361/201833559},
  \href {http://adsabs.harvard.edu/abs/2019A%26A...622A...1S} {622, A1}

\bibitem[\protect\citeauthoryear{{Slee}}{{Slee}}{1995}]{Slee95}
{Slee} O.~B.,  1995, \mn@doi [Australian Journal of Physics]
  {10.1071/PH950143}, \href
  {https://ui.adsabs.harvard.edu/abs/1995AuJPh..48..143S} {48, 143}

\bibitem[\protect\citeauthoryear{{Smirnov} \& {Tasse}}{{Smirnov} \&
  {Tasse}}{2015}]{Smirnov+Tasse15}
{Smirnov} O.~M.,  {Tasse} C.,  2015, \mn@doi [\mnras] {10.1093/mnras/stv418},
  \href {http://adsabs.harvard.edu/abs/2015MNRAS.449.2668S} {449, 2668}

\bibitem[\protect\citeauthoryear{{Tasse}}{{Tasse}}{2014}]{Tasse14}
{Tasse} C.,  2014, \mn@doi [\aap] {10.1051/0004-6361/201423503}, \href
  {http://adsabs.harvard.edu/abs/2014A%26A...566A.127T} {566, A127}

\bibitem[\protect\citeauthoryear{{Tasse} et~al.,}{{Tasse}
  et~al.}{2018}]{Tasse+18}
{Tasse} C.,  et~al., 2018, \mn@doi [\aap] {10.1051/0004-6361/201731474}, \href
  {http://adsabs.harvard.edu/abs/2018A%26A...611A..87T} {611, A87}

\bibitem[\protect\citeauthoryear{{Werner}, {Worrall}  \& {Birkinshaw}}{{Werner}
  et~al.}{1999}]{Werner+99}
{Werner} P.~N.,  {Worrall} D.~M.,   {Birkinshaw} M.,  1999, \mn@doi [\mnras]
  {10.1046/j.1365-8711.1999.02677.x}, \href
  {https://ui.adsabs.harvard.edu/\#abs/1999MNRAS.307..722W} {307, 722}

\bibitem[\protect\citeauthoryear{{Wirth}, {Smarr}  \& {Gallagher}}{{Wirth}
  et~al.}{1982}]{Wirth+82}
{Wirth} A.,  {Smarr} L.,   {Gallagher} J.~S.,  1982, \mn@doi [\aj]
  {10.1086/113135}, \href
  {https://ui.adsabs.harvard.edu/\#abs/1982AJ.....87..602W} {87, 401}

\bibitem[\protect\citeauthoryear{{Worrall}, {Birkinshaw}  \&
  {Cameron}}{{Worrall} et~al.}{1995}]{Worrall+95}
{Worrall} D.~M.,  {Birkinshaw} M.,   {Cameron} R.~A.,  1995, \apj, 449, 93

\bibitem[\protect\citeauthoryear{{Wyithe} \& {Loeb}}{{Wyithe} \&
  {Loeb}}{2003}]{Wyithe+03}
{Wyithe} J. S.~B.,  {Loeb} A.,  2003, \mn@doi [\apj] {10.1086/375187}, \href
  {https://ui.adsabs.harvard.edu/\#abs/2003ApJ...590..691W} {590, 691}

\bibitem[\protect\citeauthoryear{{Zhang}, {Dultzin-Hacyan}  \& {Wang}}{{Zhang}
  et~al.}{2007}]{Zhang+07}
{Zhang} X.-G.,  {Dultzin-Hacyan} D.,   {Wang} T.-G.,  2007, \mn@doi [\mnras]
  {10.1111/j.1365-2966.2007.11673.x}, \href
  {http://adsabs.harvard.edu/abs/2007MNRAS.377.1215Z} {377, 1215}

\bibitem[\protect\citeauthoryear{{Zier}}{{Zier}}{2005}]{Zier05}
{Zier} C.,  2005, \mn@doi [\mnras] {10.1111/j.1365-2966.2005.09586.x}, \href
  {https://ui.adsabs.harvard.edu/abs/2005MNRAS.364..583Z} {364, 583}

\bibitem[\protect\citeauthoryear{{Zwicky} \& {Kowal}}{{Zwicky} \&
  {Kowal}}{1968}]{Zwicky+Kowal68}
{Zwicky} F.,  {Kowal} C.~T.,  1968, {''Catalogue of Galaxies and of Clusters of
  Galaxies'', Volume VI}

\bibitem[\protect\citeauthoryear{{van Haarlem} et~al.,}{{van Haarlem}
  et~al.}{2013}]{vanHaarlem+13}
{van Haarlem} M.~P.,  et~al., 2013, \mn@doi [\aap]
  {10.1051/0004-6361/201220873}, \href
  {http://adsabs.harvard.edu/abs/2013A%26A...556A...2V} {556, A2}

\bibitem[\protect\citeauthoryear{{van Weeren}, {de Gasperin}, {Akamatsu},
  {Br{\"u}ggen}, {Feretti}, {Kang}, {Stroe}  \& {Zandanel}}{{van Weeren}
  et~al.}{2019}]{vanWeeren+19}
{van Weeren} R.~J.,  {de Gasperin} F.,  {Akamatsu} H.,  {Br{\"u}ggen} M.,
  {Feretti} L.,  {Kang} H.,  {Stroe} A.,   {Zandanel} F.,  2019, \mn@doi [\ssr]
  {10.1007/s11214-019-0584-z}, \href
  {http://adsabs.harvard.edu/abs/2019SSRv..215...16V} {215, 16}

\makeatother
\end{thebibliography}

\end{document}